# Synchronized and Desynchronized Phases of Exciton-Polariton Condensates in the Presence of Disorder

A. Baas, K. G. Lagoudakis, M. Richard, R. André[†], Le Si Dang[†], B. Deveaud-Plédran

IPEQ, Ecole Polytechnique Fédérale de Lausanne (EPFL), Station 3, CH-1015 Lausanne, Switzerland,
[†] Institut Néel, CNRS, 25 Avenue des Martyrs, 38042 Grenoble, France

*Condensation of exciton-polaritons in semiconductor microcavities takes place despite in plane disorder. Below the critical density the inhomogeneity of the potential seen by the polaritons strongly limits the spatial extension of the ground state. Above the critical density, in presence of weak disorder, this limitation is spontaneously overcome by the non linear interaction, resulting in an extended synchronized phase. This mechanism is clearly evidenced by spatial and spectral studies, coupled to interferometric measurements. In case of strong disorder, several non phase-locked (independent) condensates can be evidenced. The transition from synchronized phase to desynchronized phase is addressed considering multiple realizations of the disorder.*

Quantum phase transitions realized in inhomogeneous potentials is a topic of major importance in condensed matter physics. It involves non trivial matter states, as Josephson oscillations for double potential wells[1], Mott insulator for periodic potentials[2], and Bose glass for disorder potentials[3]. The case of disorder potentials is a crucial matter both for the intrinsic interest of phase transitions and for the unavoidable disorder in real systems[4]. The recent achievement of Bose Einstein condensation for microcavity polaritons [5] addresses again this issue. Polariton condensation indeed takes place in a disordered medium that creates traps, in which the polaritons are preferentially located. Polaritons are two-dimensional quasi-particles, resulting from the strong coupling of cavity photons with quantum well excitons. Their very low mass allows the observation of condensation at unexpectedly high temperatures. Due to their short radiative recombination time, the phase transition gets a non-equilibrium character[6]. The fundamental difference with equilibrium phase transitions is that here the stationary state corresponds to a balanced flux of incoming and outgoing particles in the condensate. The cw pump laser intensity, that determines the number of particles in the system, plays the role of the driving parameter of the phase transition. Due to their excitonic part, polaritons can interact. At the densities necessary to achieve condensation, these interactions happen in the dilute limit[7], and are evidenced through the spectral blue shift of the states. To these two intrinsic features of such light-matter quasi particles adds a particularity of the II-VI semiconductor microcavity that displays cavity length fluctuations[8]. This results in strong fluctuations of the polariton energies over small distances, which create some sort of traps for polaritons. Eventually, the nature of the condensed phase will be the result of the complex interplay between disorder effects, linear coupling and non-linear interactions within the very short lifetime of the polaritons.

In this Letter we focus on the role of disorder on the spontaneous formation of a two dimensional extended polariton condensate. This issue differs from the one discussed in atomic systems[9], where the condensate is created in the absence of disorder and eventually experiences a disordered potential. At low densities, the initial polariton cloud is already subject to the in-plane disorder, which is responsible for spectral fluctuations of the polariton states. We shall give experimental evidence that at high polariton density, non linear interactions - manifested by the polariton blue shift- are necessary for the formation of a synchronized phase at high intensities, i.e. the macroscopic occupancy of a single coherent quantum state - delocalized over several potential traps - not originally existing. This site to site synchronization is demonstrated by coupled spectral and interferometric measurements. To our knowledge, this is the first experimental evidence of such a frequency locking for a polaritonic system undergoing a phase transition. For some configurations of the disorder depending on the position on the sample, a desynchronized phase with several non synchronized condensates is reached, corresponding to a non uniform order parameter.

Synchronization is a general behavior observed in various physical systems since the historical Huygens coupled pendulums[10]: arrays of Josephson oscillators[11], arrays of VCSEL's[12], or the microwave spin torque nano-oscillators[13]. The basic idea of this effect is that initially independent oscillators end up sharing a correlated evolution, i.e. they are linked by some phase relation, whether their frequencies are equal or not. This can be achieved by an internal or an external force, resulting in scattering or injecting some portion of one oscillator into its neighbors. The oscillators participating to a synchronized system must possess a non linear response to force synchronization.

Our study of the condensate formation reveals two different cases, one for which a synchronized phase is reached despite disorder and one for which non linear interactions don't allow to overcome the existing disorder. These two different behaviors will be highlighted by a set of experiments implemented at two different positions on the same sample (Pos.1 and Pos.2, see Figure 1). These positions only differ by the local potential disorder seen by the polaritons.

The sample used for our experiments is the same one as in our previous works[8],[5]. The excitation is performed in a non resonant quasi-CW way. Spatial and spectral resolutions are achieved using a real space imaging setup in conjunction with a 10μeV resolution spectrometer. The diffraction limited real space image of the sample surface (Figure 1) is formed on the entrance slits of the spectrometer. This allows us to select one line (x=0) in real space and to resolve spectrally the luminescence from each point of this line. On Figure 2 the vertical axis corresponds to the position along a line of the sample that is selected by the spectrometer and the horizontal axis to the energy. This image contains no information on the angular emission from the microcavity since the emitted light is integrated from -30 to 30 degrees. Let us stress that the ground state (that emits light at normal incidence) and all the excited states (that emit light in the selected cone) contribute to the measured luminescence.

The integrated luminescence emitted by any area of the sample appears to be homogeneous for low pump intensities [8],[5]. This changes dramatically when we spectrally resolve the signal. As shown in Figure 2(a) for Pos. 1, we observe clear fluctuations of the ground state energy, which are due to the disorder of the sample structure. At low densities, we clearly observe a non thermal distribution with a bottleneck effect on the lower polariton branch dispersion image (not shown). When the pump power is increased towards the condensation threshold, the bottleneck disappears and the polariton cloud becomes thermalized at 19.2 K due to increased polariton-polariton scattering. As the pump is increased above threshold, two pronounced effects are evidenced at Pos.1: (i) there is a common blue shift across the whole excitation region; (ii) a spectrally homogeneous delocalized ground state appears over the entire excitation region (see Figure 2(b)).

To demonstrate unambiguously that this single ground state is only defined above threshold, we put in evidence the first order correlations by means of a modified Michelson interferometer[5] with spectral resolution along the line studied in the previous experiment. The region of significant contrast away from the central peak of autocorrelation in Figure 2(c-d) shows the spatial extension (vertical axis) of the ground state. At higher energy E($|\mathbf{k}|$), all the states of the same wave vector **k** in modulus contribute, and correlations vanish. Below threshold (Figure 2(c)), the spatial extension of the first order coherence is limited by disorder to a few microns: there is no common ground state for polaritons localized in different potential minima. This experiment rules out any description exclusively formulated in terms of linear coupling, which should define one single ground state irrespective of the excitation density. Above threshold the coherent domain reaches the size of the area excited by the laser in case of Pos. 1 (this enhancement is reduced in case of Pos. 2, see below).

We investigate the evolution of the frequency detuning of spatially separated luminescent spots of the condensate (see Figure 1(a)) across the condensation transition. The energy of the ground state (indicated by a small vertical line on Figure (a,b,f,e)) is obtained using a lorentzian fit on the low energy side, with a fixed linewidth corresponding to the ground state (350 μeV, measured separately). The frequency detuning is then defined as the frequency difference between the energies of the fitted lorentzians at the two different positions. We selected four luminescent points (see Figure 1(a)) and monitored the frequency detuning for the different pairs of neighbors (see Figure 3). When increasing the density over the critical value, we observe a very clear reduction/collapse of the frequency detuning for all pairs of neighbors: The initial detunings are of the order of 300 μeV below threshold and become smaller than 75μeV above threshold. This residual detuning is very small compared to the linewidth of the condensate (at minimum 400 μeV). It is also negligible when compared to the global blue shift, which is 500 μeV between $P_{th}$ and $3P_{th}$.

In order to relate this transition with the formation of a polariton condensate, we have to consider again the spatial coherence properties. We concentrate here on a *two dimensional* mapping of $g^{(1)}(\mathbf{r},-\mathbf{r})$ [5]. The region over which correlations are significant, defines the condensate domain. For Pos. 1, it demonstrates that all the points for which we observe a clear reduction/collapse of the frequency detuning belong to the same condensate, a feature marked by the contour line in Figure 1(a).

From the experiments presented above, we conclude that the formation of extended two dimensional (2D) polariton condensates involves a non linear interaction. It can come both from condensate-condensate interactions and from the reservoir of excitons and polaritons in the condensate[14]. Nonetheless this does not exclude a linear coupling such as for example photonic coupling through the potential barrier, necessary for the synchronization to occur. Our understanding is that non-linear interactions allow, via the blue shift of the states, the reduction of the initial detuning between polariton states corresponding to different potential minima, up to the point where the linear coupling can operate to induce synchronization. Experimentally, a study of the contribution of these different effects would require a full control of the disorder potential. Here we can only give the characteristics of this potential: the relative distance between neighbor potential minima varies from 2 to 6 μm, the relative difference of the potential depths shows large fluctuations up to 500 μeV, and the shape of the traps is irregular. These features explain why the possibility to get a synchronized phase depends strongly on the position on the sample[15].

We have repeated the same experiments on a different position on the sample (Pos. 2), at the same slightly positive exciton-photon detuning (between 0 and 3 meV), and the general behavior differs dramatically (see Figure (e-f)). The diamond points in Figure 3 present the evolution of the frequency detuning between polaritons at two different positions -1' and 2'- (see Figure 1(b)). They start with an initial frequency detuning of 300 μeV, which decreases slightly at the critical density but remains significant at high densities. Thus, although this frequency detuning is lower than the linewidth and the blue shift, there is no synchronization, as demonstrated in Figure 2(b). This suggests that condensed polaritons at 1' and 2' are independent. This can be evidenced through a proper mapping of the correlations, as described above. It is well known that, similarly to the case of two independent lasers[16], two independent condensates will interfere during the shortest coherence time[17]. Considering the 6 ps coherence time of the polariton condensate[5], the interference between independent polariton condensates will never be observable. Thus the lack of interference between condensed polaritons in two different potential wells can be used as a direct proof that they belong to two independent condensates (Let us note that in this case a spectrally resolved interferometric measurement is not relevant). By repeating the mapping experiment at proper positions, we were able to identify three independent condensates at Pos.2. As expected, polaritons at position 1' and 2' belong to two independent condensates. Let us stress that they have a non-negligible spectral overlap - they are separated by less than the linewidth of the condensed state-thus it is meaningful to probe if they are correlated or not, via an interferometric measurement.

We finally end up with two opposite cases for the condensation of polaritons localized in neighboring potential wells: either they merge into one single condensate, or they separate into independent condensates. In the last part of this paper, we propose to study the transition between these two cases. To do so, we choose to discriminate between dependent and independent condensates by comparing the frequency detuning in the condensed phase $\Delta\nu$ with one parameter: the initial frequency detuning $\Delta\Omega_o$. Pairs of polaritons belonging to one single condensate will appear on the $\Delta\nu=0$ line and independent condensed polaritons on the $\Delta\nu=\Delta\Omega_o$ line (Figure 4). This approach is a standard practice in the description of systems of coupled oscillators, introduced initially by Winfree[18] and Kuramoto[19] in biological systems exhibiting mode synchronization. It has been theoretically extended and applied for example to fireflies (collective flashing) and crickets (chirp in unison)[20]. This approach is also at the base of the Adler equation[21] and can be used to describe laser mode locking[22]. Figure 4 presents the results of a statistical study over a large region of zero exciton-photon detuning on the sample. The general behavior corresponds very well to what is expected for the two cases. The points appearing between the two limit cases of synchronized and desynchronized phase correspond to pairs of polaritons that do not feel the non linear interaction intensely enough to get synchronization. This transition can be described in the frame of the Adler equation, where only a non linear coupling is taken into account. Considering two coupled oscillators with initial frequency detuning $\Delta\Omega_o$, the evolution of the relative phase $\psi(t)$ between them is: $d\psi(t)/dt= \Delta\Omega_o + g\sin(\psi(t))$, with g the coupling constant. Synchronization occurs when dψ(t)/dt=0, which can be achieved only if g is greater than $\Delta\Omega_o$. The transition can thus be described with only one fitting parameter, the critical value of $\Delta\Omega_o$ below which synchronization occurs, that gives an upper bound for an effective polariton coupling constant (see Fig. 4). For a model specific to polariton condensation, we refer to a very recent paper of M. Wouters[14], where a complete understanding of the conditions required to observe a synchronized phase beyond the Adler equations can be found.

To conclude, in the present work we have characterized the formation of polariton condensate(s) in a disordered medium. Synchronization allows building up a single ground state extended over a region larger than the typical disorder correlation length. From what we understand, the observation of synchronization rules out the interpretation of our results in terms of Bose glass, recently addressed for polaritons[23], where such coherence is not expected[24]. However, at some positions the disorder can destroy these correlations and produces independent condensates, clearly evidenced by our technique of correlation mapping. It could be interesting to address this question in the case of polariton parametric scattering, where a signal emission structured in several peaks has been observed[25]. This suggests lack of a synchronized phase; however as they present a non-negligible spectral overlap, the possibility of a long range phase correlations could be tested by interferometric measurements, as has been recently reported[26]. The control of the disorder appears as a crucial issue. Along this line, 0D polariton traps of any shape and size can be created in 2D microcavities[27]. On the present sample, focusing on a single potential well will allow us to minimize the influence of the disorder. This should prove useful for the characterization of the first excited modes, and the investigation of the interactions in the condensate.

**Acknowledgements**
We thank M. Wouters, I. Carusotto, D. Sarchi, V. Savona, O. El Daïf, S. Kundermann, G. Schivardi, J.-L. Staehli and B. Pietka for fruitful discussions. We acknowledge support by the Swiss National Research Foundation through the NCCR Quantum Photonics.

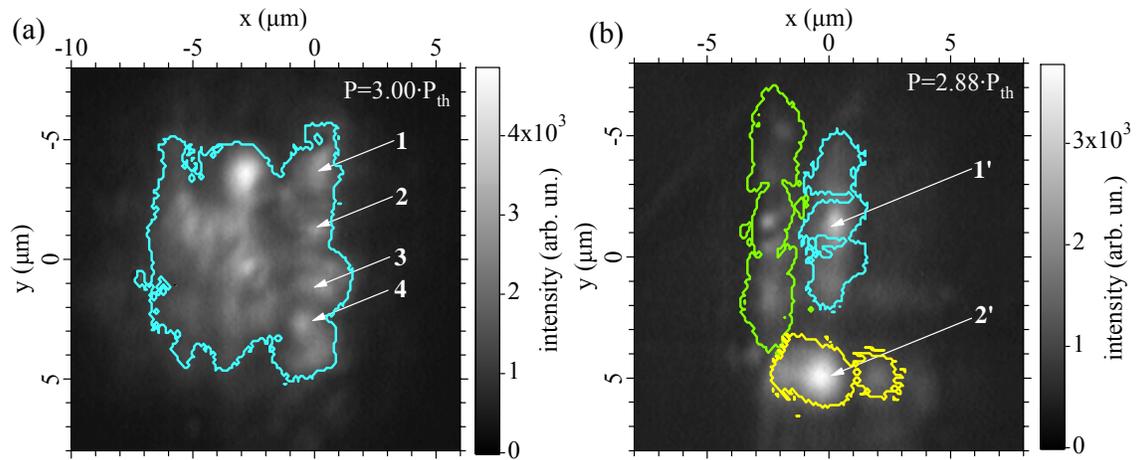

Figure 1. (Color online) (a) Real space luminescence of one single condensate taken at Pos.1 and (b) three independent condensates, taken at Pos.2. The excitation conditions and the detuning are the same for both.

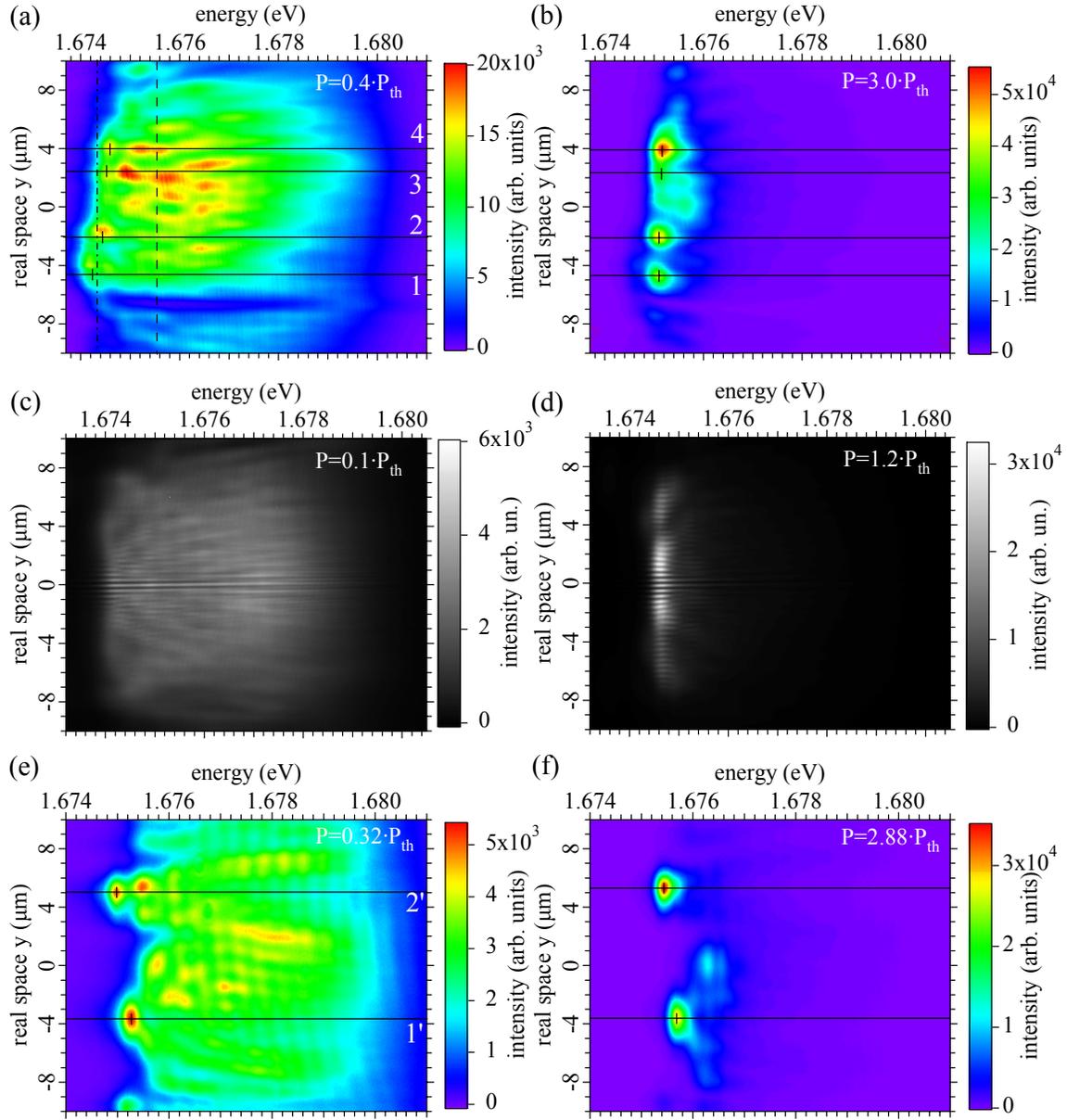

Figure 2. (Color online) (a-b and e-f) Spectrally resolved real space line and (c-d) interferogram. Horizontal axis: energy in electron volts. Vertical axis: real space in microns. (a) profile along x=0 line of Fig. 1(a) (Pos.1), at an excitation density below threshold P=0.4·$P_{thr}$. At the same position, the ground state energy given by the spatially integrated dispersion curve is at 1.67433 eV (dashed-dotted vertical line). The bottleneck appears on the same dispersion curve around 1.67554 eV (dashed line). Small vertical lines define the energy of the ground state calculated by a lorentzian fit at the low energy side (b) same above threshold P=3·$P_{thr}$. (c) Spectrally resolved interferogram below threshold P=0.1·$P_{thr}$ and (d) same above threshold P=1.2·$P_{thr}$. (e) profile along x=0 line of Fig. 1(b) (Pos.2), at an excitation density below threshold P=0.32·$P_{thr}$. (f) same above threshold P=2.88·$P_{thr}$.

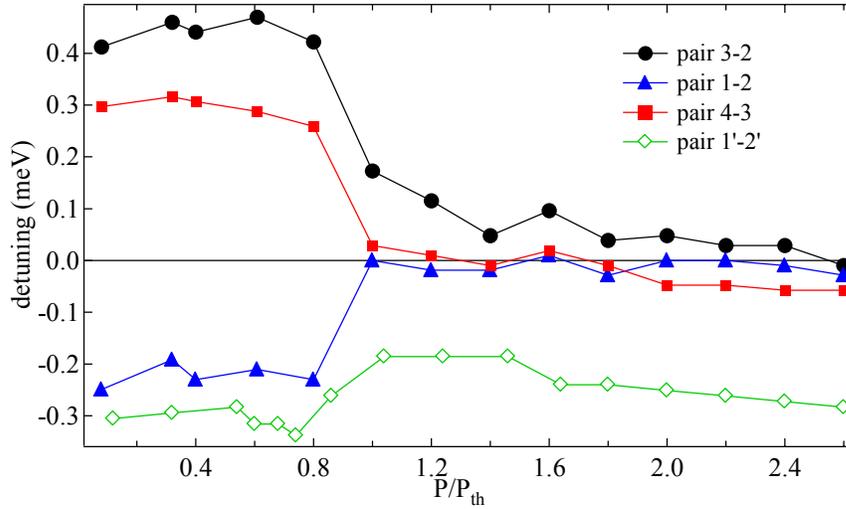

Figure 3. (Color online) Frequency detuning for different pairs of spatially separated condensed polaritons as a function of the excitation intensity. Vertical axis: frequency detuning in meV. Horizontal axis: power normalized to the threshold power of condensation $P_{thr}$. The experimental points are connected by lines. Points 1, 2, 3 and 4 belong to a single condensate (Pos. 1) and points 1' and 2' belong to independent condensates (Pos. 2).

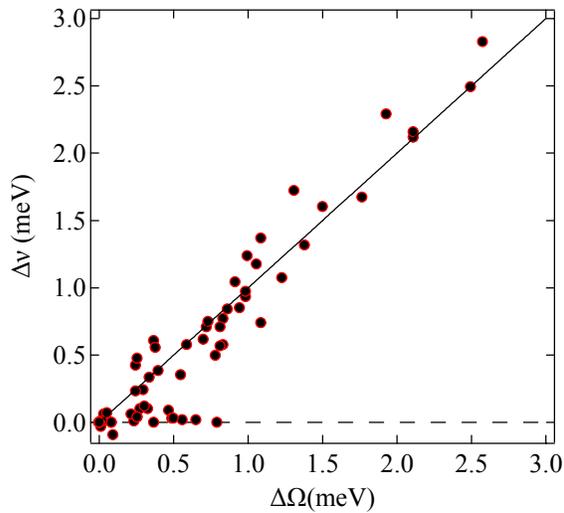

Figure 4. (Color online) Statistical study of the frequency detuning above threshold ($\Delta\nu$) as a function of the frequency detuning well below threshold ($\Delta\Omega_o$). Solid line represents the non synchronized phase: $\Delta\nu = \Delta\Omega_o$, dashed line represents the synchronized phase: $\Delta\nu = 0$.

11